\documentclass[aps,prb,twocolumn,superscriptaddress,showpacs]{revtex4}

\usepackage[dvips]{epsfig}
\usepackage[dvips]{color}
\usepackage{amssymb}

\newcommand{\ytio}{Y$_2$Ti$_2$O$_7${}}
\newcommand{\gdtio}{Gd$_2$Ti$_2$O$_7${}}
\newcommand{\gdsno}{Gd$_2$Sn$_2$O$_7${}}
\newcommand{\ygd}{(Y$_{0.995}$Gd$_{0.005}$)$_2$Ti$_2$O$_7$ {}}
\newcommand{\Stev}[2]{\hat{O}^{#1}_{#2}}
\newcommand{\BStev}[2]{B^{#1}_{#2}\hat{O}^{#1}_{#2}}
\newcommand{\eqnref}[1]{(\ref{#1})}
\begin{document}

\title{Single-ion anisotropy in the gadolinium pyrochlores studied
by an electron paramagnetic resonance} 

\author{V.~N.~Glazkov}
\affiliation{Commissariat \`a l'Energie Atomique, DSM/DRFMC/SPSMS,
38054 Grenoble, Cedex 9, France}
\affiliation{P.~L.~Kapitza Institute for Physical Problems RAS, 117334
 Moscow, Russia}

\author{M.~E.~Zhitomirsky}
\affiliation{Commissariat \`a l'Energie Atomique, DSM/DRFMC/SPSMS,
38054 Grenoble, Cedex 9, France}

\author{A.~I.~Smirnov}
\affiliation{P.~L.~Kapitza Institute for Physical Problems RAS, 117334
 Moscow, Russia}

\author{H.-A.~Krug von Nidda}

\affiliation{Experimentalphysik V, EKM, Institut f\"{u}r Physik,
Universit\"{a}t Augsburg, 86135 Augsburg, Germany}

\author{A.~Loidl}

\affiliation{Experimentalphysik V, EKM, Institut f\"{u}r Physik,
Universit\"{a}t Augsburg, 86135 Augsburg, Germany}

\author{C. Marin}
\affiliation{Commissariat \`a l'Energie Atomique, DSM/DRFMC/SPSMS,
38054 Grenoble, Cedex 9, France}

\author{J.-P. Sanchez}
\affiliation{Commissariat \`a l'Energie Atomique, DSM/DRFMC/SPSMS,
38054 Grenoble, Cedex 9, France}

\date{\today}

\begin{abstract}
The electron paramagnetic resonance is used to measure the single-ion
anisotropy of Gd$^{3+}$ ions in the pyrochlore structure of
(Y$_{1-x}$Gd$_x$)$_2$Ti$_2$O$_7$ . A rather strong easy-plane type
anisotropy is found. The anisotropy constant $D$ is comparable to the
exchange integral $J$ in the prototype Gd$_2$Ti$_2$O$_7$, $D\simeq
0.75J$, and exceeds the dipolar energy scale. Physical implications
of an easy-plane anisotropy for a pyrochlore antiferromagnet are
considered. We calculate the magnetization curves at $T=0$ and
discuss phase transitions in magnetic field.
\end{abstract}

\maketitle

Pyrochlore antiferromagnets have been actively studied in the past due
to their unusual properties.\cite{hfm03}
The magnetic ions in these compounds form a network of corner-sharing
tetrahedra, which is prone to a high degree of geometric
frustration. The ground state of a classical Heisenberg
antiferromagnet on a pyrochlore lattice is macroscopically
degenerate and remains disordered.\cite{classical}
Since weaker residual interactions are always present
in real magnetic materials, an important question is
about a separation of energy scales. The corresponding knowledge
helps to clarify how the macroscopic entropy freezes
as $T\to 0$ and what type of magnetically ordered, spin-liquid,
or glassy state is eventually stabilized.

Gadolinium pyrochlores Gd$_2$Ti$_2$O$_7$ and Gd$_2$Sn$_2$O$_7$ are
considered to be the best realizations of the Heisenberg model on a
pyrochlore lattice, since Gd$^{3+}$ is, nominally,  in a $^8S_{7/2}$
state with completely frozen orbital degrees of freedom. Both
compounds order at temperatures of about 1~K,
\cite{raju,ramirez,bonville} while Gd$_2$Ti$_2$O$_7$  exhibits also
multiple phases in magnetic field. \cite{ramirez,petrenko} This has
been attributed to weaker dipolar interactions between the $S=7/2$
gadolinium spins,\cite{raju,palmer,cepas} though no unambiguous
explanation neither of the phase diagram of Gd$_2$Ti$_2$O$_7$ nor of
the neutron-diffraction measurements \cite{stewart} has been
presented.

Single-ion effects have so far been neglected for the two gadolinium
pyrochlores. However, strong spin-orbit coupling of the $4f$
electrons breaks the $LS$-scheme of the energy levels of the
Gd$^{3+}$ ion and mixes the $^8S_{7/2}$ and $^6P_{7/2}$
states.\cite{antic-fidancev} Since the ground-state multiplet
contains an admixture of the $L\neq 0$ states, it can be split by the
crystal field. Crystal-field splitting of the order of $0.8$~K was
observed for Gd$^{3+}$ ions in yttrium-gallium garnet. \cite{garnet}
The amplitude of such a splitting is comparable to the exchange
coupling and the dipolar energy in the prototype
Gd$_3$Ga$_5$O$_{12}$. Thus, an accurate study of the single-ion
anisotropy in the Gd-based pyrochlores is necessary to understand the
magnetic properties of these materials.

To study properties of isolated Gd$^{3+}$ ions we have prepared
single-crystals of non-magnetic \ytio{} diluted with a small amount
(nominally 0.5\%) of Gd. The crystals were grown by traveling solvent
floating zone technique. After growth the oriented single-crystals
were annealed in an oxygen atmosphere. The absence of parasitic
phases has been verified by powder X-ray diffraction. The
magnetization of the reference sample of nominally pure \ytio{}
corresponds to no more than 0.06\% of $S=1/2$ impurities per yttrium.
The lattice parameters and geometry of the oxygen surrounding of the
rare-earth site are very close for  \ytio{} and \gdtio{}.
\cite{structure} Hence, single-ion parameters for Gd$^{3+}$ ions in
\ygd{} and \gdtio{} are also expected to be very close.  An electric
field gradient at the Gd site in \gdsno{} determined from the
quadrupolar  M\"ossbauer spectra\cite{bonville} is 30\% smaller than
that of \gdtio{}, which should lead to the corresponding decrease of
the single-ion anisotropy.

Electron paramagnetic resonance (EPR) is a conventional tool
to probe small crystal-field splitting of  the energy
levels of magnetic ions. \cite{abragham}
The zero-field splitting of the multiplet leads to an
easily detected change of the resonant absorption frequencies. We have used
a set of home-made spectrometers with transmission type cavities for
the high-frequency (20--115~GHz) measurements, and a commercial
Bruker EPR spectrometer for the high-precision X-band (9.36~GHz)
and Q-band (34.0~GHz) measurements.

The local surrounding of the rare-earth site in the pyrochlore structure
corresponds to a trigonal point symmetry  $D_{3d}$ ($\overline{3}m$).
A general form of the single-site Hamiltonian
is, then,
\begin{eqnarray}
\hat{\cal H} & = & \BStev{0}{2}+\BStev{0}{4}+\BStev{3}{4}+\BStev{0}{6}
\nonumber\\
&& \mbox{} +\BStev{3}{6}+\BStev{6}{6}-g\mu_B\mathbf{H}\cdot\hat{\mathbf{S}}
\ ,
\label{Hcf}
\end{eqnarray}
where the Stevens operators \cite{abragham,T-curie-weiss} $\Stev{i}{j}$ are functions
of the components of the {\em total} angular momentum $S=7/2$, with
$\hat{\bf z}\parallel\langle111\rangle$ and $\hat{\bf x}\parallel\langle
11\overline{2}\rangle$.
The lowest-order anisotropy
term $\BStev{0}{2}$ can be rewritten
up to an additive constant
as $DS^2_z$ with $D=3B^0_2$.

\begin{figure}
\centering
\epsfig{file=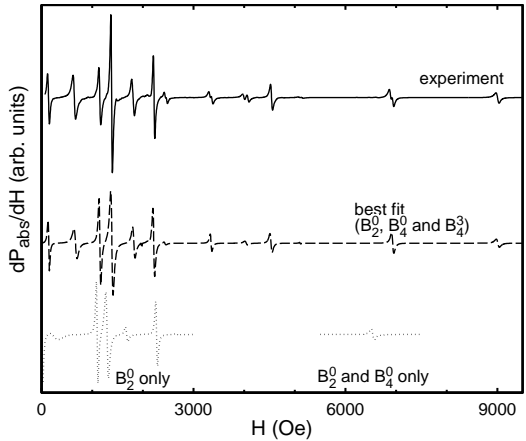, width=0.8\columnwidth}
\caption{Observed derivative of the 9.36~GHz microwave absorption (solid);
modeled curve best fitting to the experimental results (dashed);
parts of the modeled curves (dotted)
corresponding to the spin-Hamiltonian constricted to the second-order term
only (left) and second- and fourth-order diagonal terms (right).
f=9.36~GHz, $\mathbf{H}||$[110], T=6.0~K.}
\label{fig:x110}
\end{figure}

\begin{figure}
\centering
\epsfig{file=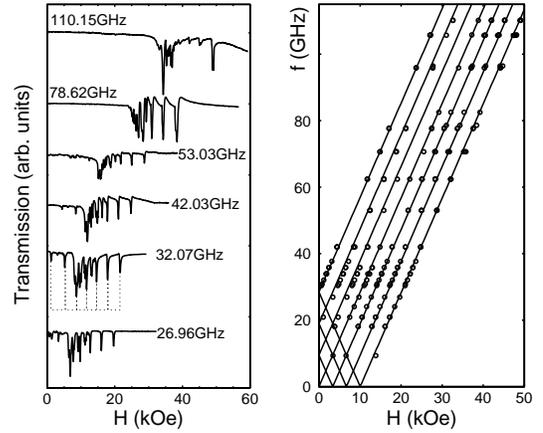, width=0.8\columnwidth}
\caption{ Left panel: Field dependence of the microwave absorption for the high-frequency EPR spectra at T=4.2K and $\mathbf{H}||[111]$. Dotted lines ($f=32.07$GHz) mark positions of the absorption determined by the spins with the anisotropy axis parallel to the field direction.
Right panel: Field dependence of the selected resonance frequencies for the spins with the anisotropy axis parallel to the field direction ($[111]$). Solid lines give theoretical fit.}
\label{fig:f(h)111}
\end{figure}

In the absence of the single-ion anisotropy Hamiltonian \eqnref{Hcf}
contains only the Zeeman term. In this case the frequencies of all
allowed transitions are equal and only a single resonance absorption
line should be observed at a field determined by
$\hbar\omega=g\mu_BH$. The measured EPR absorption spectra, see
Figs.~\ref{fig:x110} and \ref{fig:f(h)111}, contain multiple
components demonstrating the presence of a non-negligible anisotropy.
The analysis of the EPR spectra is complicated by the existence of
four inequivalent magnetic ion positions. At high frequencies and
$\mathbf{H}\parallel [111]$ it is possible to separate seven
absorption components, which correspond to the  spins with anisotropy
axis parallel to the magnetic field direction. They include
absorption component with the largest value of the resonance field
and are almost equidistant (see Fig.~\ref{fig:f(h)111}). This
indicates that the second order axial term ($DS_z^2$) is the most
important one, its magnitude estimated from the total splitting of
the spectra is $|D|\sim0.25$K. In order to refine the values of the
anisotropy constants, we have modeled numerically the EPR
absorption\cite{abragham} for different directions of magnetic field
and different microwave frequencies trying to obtain best
correspondence of the observed and modeled resonance fields values
(see, e.g., Fig.~\ref{fig:x110}). To determine the sign of the main
anisotropy constant (resonance-field position determine only relative
signs of the anisotropy constants), we have studied the temperature
dependence of the intensity of different components at high microwave
frequency. This analysis yielded that the  main anisotropy term is of
the easy-plane type and the spin Hamiltonian parameters are
$g=1.987$, $B^0_2=(74.3\pm0.3)10^{-3}$K, $B^0_4=(5.7\pm0.9)10^{-5}$K,
$B^3_4=(1.8\pm0.2)10^{-3}$K. The determination of the 6-th order
anisotropy parameters has been impossible due to a presence of small
($0.5^\circ$ or less) misorientation of the sample. We were able to
estimate only the upper bounds: $|B^0_6|<10^{-6}$K,
$|B^3_6|<2\cdot10^{-5}$K, $|B^6_6|<10^{-6}$K.

The measured value of the $B_2^0$ coefficient indicates that
the single-ion anisotropy  is the second largest interaction
in the effective spin Hamiltonian of Gd$_2$Ti$_2$O$_7$.
Indeed, the nearest-neighbor exchange integral is estimated
from the Curie-Weiss temperature \cite{raju,ramirez}
$\theta_{CW}=-9.5$~K as $JS^2\approx 3.7$~K, whereas $DS^2\approx 2.73$~K.
The dipolar constant $E_{dd} = (g\mu_BS)^2/(a\sqrt{2}/4)^3$ for the interaction
between adjacent magnetic ions for above values of the $g$-factor and
lattice parameter amounts to only $E_{dd}\approx0.65$~K.
In the following, we discuss several properties of magnetic pyrochlore
materials, which might be affected by a strong single-ion anisotropy of
the easy-plane type.

The first question to be addressed theoretically is whether
a significant single-ion anisotropy
modifies the value of the exchange constant obtained from $\theta_{CW}$.
Standard high-temperature expansion\cite{T-curie-weiss} of the susceptibility tensor
includes in general a contribution
from the single-ion term
$D\sum_i ({\bf n}_i\cdot{\bf S}_i)^2$:
\begin{eqnarray}
\chi^{\alpha\beta} & = &
\frac{g^2\mu_B^2S(S+1)}{3T}
\Bigl\{\,\delta^{\alpha\beta}
 - \delta^{\alpha\beta} \frac{S(S+1)}{3T}\sum_j J_{ij}
\nonumber \\
&& - \frac{2D}{5T}(S-\textstyle{\frac{1}{2}})(S+\textstyle{\frac{3}{2}})
\bigl(\langle n^\alpha n^\beta\rangle-\frac{1}{3}\delta^{\alpha\beta}\bigr)
\Bigr\}.
\label{chi}
\end{eqnarray}
In the case of a pyrochlore lattice the averaging in Eq.~(\ref{chi}) has
to be done over four inequivalent directions of the local anisotropy axis
${\bf n}_i$ corresponding to the principal cube diagonals. Using
the identity $\frac{1}{4}\sum_{i=1}^4 n_i^\alpha n_i^\beta
= \frac{1}{3}\delta^{\alpha\beta}$ we find that the single-ion
contribution vanishes in the second-order of the $1/T$ expansion. Therefore,
the Curie-Weiss constant is not affected by the single-ion term
and is given by the same expression as in a pure Heisenberg case:
$\theta_{CW} = zJS(S+1)/3$, where $z=6$ is the number of nearest-neighbor
magnetic atoms.

The unusual ordering wave-vector
$\bigl(\frac{1}{2}\frac{1}{2}\frac{1}{2}\bigr)$ in \gdtio{} is
probably determined by a delicate balance between the long-range
dipolar interaction and several weak next-neighbor exchanges.
\cite{cepas} The full analysis of the role of the single-ion
anisotropy for such models lies beyond the scope of present study. We
will  discuss instead properties of a simplified classical spin model
on a pyrochlore lattice, which takes into account only the
nearest-neighbor exchange interaction and a staggered single-ion
term:
\begin{equation}
\hat{\cal H} = J \sum_{\langle ij\rangle} {\bf S}_i\cdot {\bf S}_j + D
\sum_i ({\bf n}_i\cdot {\bf S}_i)^2 - {\bf H} \cdot \sum_i {\bf S}_i \ .
\label{Hclass}
\end{equation}
Transformation to classical spins of unit length $|{\bf S}_i|=1$
amounts to the following redefinition of the physical parameters:
$J\to JS^2$, $D\to DS^2$, $H\to g\mu_BHS$. The minimum of the
Heisenberg interaction in zero field is attained for states with zero
total spin on every tetrahedron: ${\bf S}_{\rm tet}=\sum_{i=1}^4 {\bf
S}_i=0$.\cite{classical} The key observation for the anisotropic
model (\ref{Hclass}) is that the ${\bf S}_{\rm tet}=0$ constraint is
compatible with local spins being oriented perpendicular to their
local hard axes: the exchange and the anisotropy terms can be
simultaneously minimized. \cite{bramwell94,champion04} In particular,
for one tetrahedron there is one free continuous parameter (angle),
which describes the possible ground states. The remaining degeneracy
of the lattice model is infinite but not extensive: the number of
continuous parameters describing the classical ground state scales as
$O(L^2)$ with the linear system size $L$. Monte Carlo investigations
\cite{bramwell94,champion04} of the model (\ref{Hclass}) with $D/J=5$
and $\infty$  have found a first order transition into a noncoplanar
`$q=0$' state at $T_c\simeq 0.13J$, which is driven by the thermal
order by disorder mechanism.

To complement the above zero-field consideration we now discuss the
magnetization process of the model (\ref{Hclass}). The total spin
operator does not commute with the Hamiltonian in the presence of
anisotropic interactions. Consequently, the magnetization plateaus,
{\it e.g.\/}, the 1/2-plateau, do not appear on the magnetization
curves of the gadolinium pyrochlores and $M(H)$ approaches the
saturation value at high fields only asymptotically. Besides, the
staggered (within one unit cell) single ion anisotropy might also
smear phase transitions in finite magnetic fields for `nonsymmetric'
field orientations.

\begin{figure}
\centering \epsfig{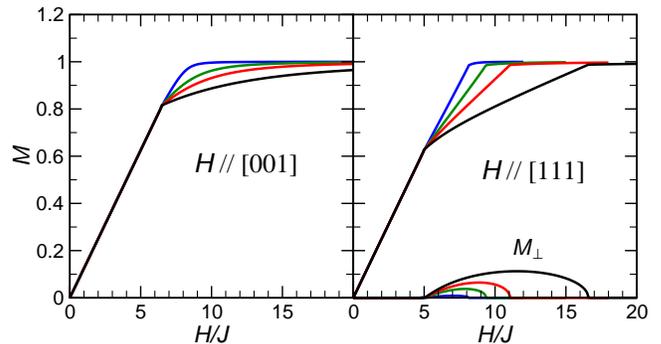}
\caption{(Color online) Magnetization per spin for the
 classical easy-plane pyrochlore antiferromagnet at $T=0$
 for the field applied along the two symmetry directions.
 Magnetization curves from top to bottom are drawn for
 $D/J=0.2,1,2$, and 5. The transverse magnetization
 on the right panel corresponds to the same choice of $D/J$ in the
 reverse order. }
\label{Mlong}
\end{figure}

Returning back to the Hamiltonian (\ref{Hclass}) the minimum of
the Heisenberg and the Zeeman interactions is given by the states
obeying ${\bf S}_{\rm tet} ={\bf H}/2J$. \cite{mzh}
Simple extension of the zero-field consideration suggests that in weak magnetic
fields spins remain in their respective local easy-planes. It is again
possible to simultaneously minimize the local single-ion anisotropy terms
and to satisfy the magnetization constraint. The magnetization
curve is initially a straight line: $M(H)=H/8J$. The remaining degeneracy
corresponds to one continuous degree of freedom per tetrahedron and $O(L^2)$
variables for the lattice problem.

The situation, however, changes with increasing magnetic field. At a
certain critical field $H_c$, which is smaller than the saturation
field of the Heisenberg model $H_{\rm sat}=8JS$, the constraint ${\bf
S}_{\rm tet} ={\bf H}/2J$ cannot be satisfied anymore by spins lying
in their local easy-planes. The degeneracy is lost and spins start to
tilt out of the easy-planes towards the field direction. The
corresponding critical field does not depend on the value of the
anisotropy constant and is determined purely by geometry. It can be
calculated analytically for fields applied along the symmetry
directions  of a pyrochlore lattice:
\begin{eqnarray}
&& H_c^{[001]}=8J\sqrt{2/3}\ ,  \ \
H_c^{[111]}=8J(3+\sqrt{2})/7\ , \nonumber \\
&& H_c^{[112]}=6J\ , \ \
H_c^{[110]}= 4J(1+1/\sqrt{3}) \ .
\label{Hc}
\end{eqnarray}
The fields $H_c^{[001]}$ and $H_c^{[111]}$ give, respectively, the
maximal and the minimal value for $H_c(\theta,\varphi)$. The
magnetization process at $T=0$ for the above two orientations found
by standard numerical minimization of the classical energy of a
single spin tetrahedron is presented in Fig.~\ref{Mlong}. For ${\bf
H}\parallel[001]$ the magnetization curve exhibits a kink at $H_c$
and, then, the behavior of $M(H)$ becomes nonlinear with full
saturation $M=1$ reached only asymptotically. A nondegenerate spin
structure above $H_c$ preserves the $C_2$ rotational symmetry about
the field direction, which is broken at $H<H_c$. Thus, $H_c$
describes a robust phase transition for this field orientation. The
zero-temperature magnetization curve for arbitrary field orientation
resembles the behavior for ${\bf H}\parallel[001]$. The main
difference is an absence of a spin-rotational symmetry about the
field direction, which may lead to a smearing of the phase transition
at $H=H_c$. The other two field orientations, which have stable phase
transitions at $H=H_c$ are those for the field along the $[110]$ and
the [111] axes. In the latter case the magnetization curve exhibits
two kinks: at $H_{c1}\equiv H_c\approx 5.045J$ (\ref{Hc}) and at a
second critical field $H_{c2}>H_{\rm sat}$, see the right panel in
Fig.~\ref{Mlong}. Above $H_{c2}$ the spin structure preserves the
$C_{3v}$ symmetry of the system in magnetic field, while for
$H<H_{c2}$ only a mirror symmetry remains with respect to one of the
three mirror planes passing through the [111] axis. Such a remaining
symmetry is further lost in the low-field phase at $H<H_{c1}$. The
upper transition field has an almost linear dependence on the
anisotropy constant in a wide range of values for $D\lesssim 10$:
$H_{c2}\approx 8J + 1.7D$. We, therefore, conclude, that for ${\bf
H}\parallel[111]$ there are two phase transitions in magnetic field,
which are smeared when the field is tilted away from this symmetry
direction.

\begin{figure}
\centering
\epsfig{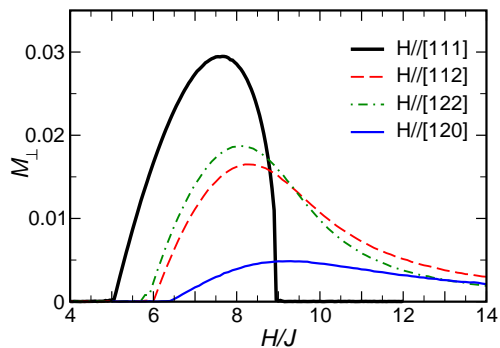}
 \caption{(Color online) Transverse magnetization per one spin of the
 classical easy-plane pyrochlore antiferromagnet at $T=0$
 for $D=0.75J$ and various field orientations.}
 \label{Mtransv}
\end{figure}

In the intermediate phase $H_{c1}<H<H_{c2}$ with broken three-fold
rotational symmetry (${\bf H}\parallel[111]$) the anisotropic
pyrochlore antiferromagnet develops a transverse magnetization, which
is shown in Fig.~\ref{Mlong}. The amplitude of the transverse
magnetization grows quickly with increasing anisotropy constant $D$.
There are, of course, three domains of ${\bf M}_\perp$ oriented under
$120^\circ$ to each other in accordance with a broken symmetry of
intermediate phase. Once magnetic field is applied away from the
$[111]$ axis, ${\bf M}_\perp$ attains a unique direction. To make
connection with Gd$_2$Ti$_2$O$_7$, which according to our
measurements has $D\approx 0.75 J$, we plot in Fig.~\ref{Mtransv} the
amplitudes of the transverse magnetization for the above ratio of
$D/J$ and for several characteristic orientations of the applied
field. Though $M_\perp$ does not exceed 3\% of the saturated
magnetization, the corresponding values can be definitely measured in
an experiment. Such measurements would also serve as a precise method
to determine the two phase transitions for ${\bf H}\parallel[111]$.
The transverse magnetization has been previously measured in
CsMnBr$_3$, a triangular lattice antiferromagnet with a strong
single-ion anisotropy. \cite{abarzhi} Note, that for ${\bf
H}\parallel[001]$ and $[110]$ the transverse magnetization does not
appear in the whole range of magnetic fields.

In conclusion, the measured local anisotropy of Gd$^{3+}$ ions
strongly affects magnetic properties of Gd-based pyrochlores. Though,
the toy model (\ref{Hclass}) is apparently not sufficient to explain
the ordered structure seen by neutrons, \cite{stewart} the quite
anisotropic phase diagram in magnetic field\cite{petrenko} is
compatible with our experimental and theoretical findings. In
particular, we have demonstrated that depending on an applied field
orientation in a {\it cubic} crystal there are possible either one or
two transitions in magnetic field. The M\"ossbauer
measurements\cite{bonville} also agree with our ESR results pointing
at ordered moments oriented perpendicular to local trigonal axes. The
presented theory of the magnetization process is relevant for
Er$_2$Ti$_2$O$_7$, which is a $D/J\to\infty$ realization of the model
(\ref{Hclass}).

The authors acknowledge support of the Russian Foundation for Basic Research
04-02-17942, of the German Bundesministerium
f\"ur Bildung und Forschung under the contract VDI/EKM
13N6917, and of the Deutsche Forschungsgemeinschaft within SFB 484 (Augsburg).

\end{document}